# Non-Stationary Time Series Model for Station Based Subway Ridership During Covid-19 Pandemic (Case Study: New York City)

**Bahman Moghimi**
Research Associate
Department of Civil Engineering
City College of New York, NY, 10031
E-mail: smoghim000@citymail.cuny.edu

**Camille Kamga**
Associate Professor
Department of Civil Engineering
City College of New York, NY, 10031
Email: ckamga@utrc2.org

**Abolfazl Safikhani**
Assistant Professor
Department of Statistics
University of Florida, Gainesville, FL
E-mail: a.safikhani@ufl.edu

**Sandeep Mudigonda**
Postdoctoral Research Associate
Region-2 University Transportation Research Center
E-mail: mudigonda@utrc2.org

**Patricio Vicuna**
Ph.D. Candidate and Research Assistant
Department of Civil Engineering
City College of New York, NY, 10031
E-mail: patricio.vicuna@gmail.com

**ABSTRACT**

The COVID-19 pandemic in 2020 has caused sudden shocks in transportation systems, specifically the subway ridership patterns in New York City. Understanding the temporal pattern of subway ridership through statistical models is crucial during such shocks. However, many existing statistical frameworks may not be a good fit to analyze the ridership data sets during the pandemic since some of the modeling assumption might be violated during this time. In this paper, utilizing change point detection procedures, we propose a piece-wise stationary time series model to capture the nonstationary structure of subway ridership. Specifically, the model consists of several independent station based autoregressive integrated moving average (ARIMA) models concatenated together at certain time points. Further, data-driven algorithms are utilized to detect the changes of ridership patterns as well as to estimate the model parameters before and during the COVID-19 pandemic. The data sets of focus are daily ridership of subway stations in New York City for randomly selected stations. Fitting the proposed model to these data sets enhances our understanding of ridership changes during external shocks, both in terms of mean (average) changes as well as the temporal correlations.

# INTRODUCTION

During outbreaks caused by infectious diseases, providing an accurate and reliable subway station ridership model is crucial for transit operators and passengers. Subway ridership changes from time to time during a year, and a statistical model that can capture this variability would enable transit authorities to plan for appropriate resource allocation, update-to-date train service frequencies for the expected ridership and sending required information to passengers during unexpected events. At the same time, with a good predict of passenger ridership, commuters can adjust their departure travels and choose other modes to reduce delays and improve their comfort.

Unforeseen and abrupt changes in urban mobility could happen due to many reasons such as crowding at the time of special events, severe weather conditions, natural disasters, infrastructure-related constructions, or global health crises like the COVID-19 pandemic. The pandemic generally reduced the overall efficiency in human mobility network, which caused an exogenous shock to the economy and is functioning similar to a natural disaster [1].

The first case of the coronavirus was confirmed in the State of New York in New York City on March 1, 2020 and since then panic buying for food and household products was reported. On March 7th, Governor Andrew Cuomo declared a state of emergency in the State of New York. A second case in the State was announced on March 10th, which was the first known case in the State to be caused through community spread. On March 16th, New York Governor Cuomo issued an executive order to close all public and private schools throughout the State, initially ordered to last until April 1st but which was later extended. On March 22nd, the New York State stay-at-home order took effect. The coronavirus cases in New York increased tremendously to the point that on March 22nd New York Times reported that the NYC became the epicenter of the pandemic [2]. As New Yorkers worked together and flatten the curve, the reopening strategies came into effect in June to let businesses to start reopening by phases. New York City entered Phase 1 of the reopening on June 8th and Phase 2 on June 22, 2020.

The coronavirus pandemic drastically altered the transportation choice behavior of all transportation users. Many travelers preferred or are obliged to stay home and work from home to comply with stay-at-home orders enacted by government officials [3]. Those with essential tasks/jobs will commute to their work location under some orders enacted to slow the spread of the virus. In order to reduce the spread of the highly contagious virus, health professionals have encouraged only essential travels. As a result, vehicular traffic volumes on roadways have been very low. Thus, the vehicular speeds in highway and corridors have increased significantly, the vehicle travel times, and delays have decreased when compared to pre-COVID time.

Public transit ridership was severely impacted by the measures enacted to slow the spread of COVID-19 viruses. In New York City, transit ridership dropped on all transit modes operated by the Metropolitan Transportation Authority (MTA) which include Subway, NYC Transit Bus, MTA Bus, Metro North Railroad, and Long Island Railroad. Specifically, for the subway ridership, the percentage change on March 12$^{th}$, March 16$^{th}$, March 23$^{rd}$, April 2$^{nd}$, April 17$^{th}$, May 15$^{th}$, and June 30$^{th}$ were reported at 19%, 60%, 87%, 92%, 93%, 90%, and 80% ridership declines respectively, as compared to similar dates in 2019 [4].

To accurately forecast transportation data and more specifically the subway ridership, one should consider such a drop in the observed time series. Many of the existing travel demand and transit ridership models are not trained to include anomalies in their models. Developing a rigorous model that can consider the ridership variability and be resilient enough to take care of anomalies is the focus of this paper. In this research, an interpretable time series model to formulate subway ridership during COVID-19 pandemic is developed. Then, statistically sound detection methods are utilized to detect the anomalies of subway ridership time series data. It is a common practice to model data with temporal index using stationary time series models, and ARIMA models are among the most well-known stationary time series models used in different scientific fields including civil engineering and transportation. However, in the presence of structural breaks (shocks) in the temporal dynamical system under consideration, the stationarity assumption may be violated; thus, one needs to search for alternative modeling frameworks. Piecewise stationary models are interesting models which are easy to interpret since a time point at which the dynamical system receives an external shock can be called a "break point", and its location can be estimated using developed algorithms. Such time points are essentially the discontinuity points in the piecewise modeling framework.

An interesting fact about the proposed detection method used in this paper is that it can handle detection for second order structure (auto-correlation). The difference between the fitted models before and after the change/break points shows that the dependence structure has changed significantly. Such changes are needed to be accounted when forecasting the ridership. This area of research is relatively undeveloped with very few literatures focused on this topic. In short, one of the main contributions of this research is to introduce a piece-wise stationary time series model which can be utilized to model sequential data experiencing external shocks. The proposed modeling framework is applied to ridership data during the covid-19 pandemic, which to the best of our knowledge, has been introduced for the first time to transportation datasets. To apply this modeling framework, a novel statistical algorithm is developed and utilized to detect break/change points, and further, fitted models before and after each break point are summarized. Certain goodness of fit tests are applied to the residuals to confirm the satisfactory performance of the proposed methodology. In the next section, a brief review of the methods found in the literature is provided.

## LITERATURE REVIEW

Subways play a crucial role in today's urban mobility. For many urban travelers, it is the convenient and first choice of mobility in large metropolitan areas. Developing a long-term transit ridership prediction is a product that may result from the conventional four-step travel demand forecasting [5]. One way to do this is to fit regression models based upon many contributing factors such as transit attributes, geographic information, demographic, and economic factors [5]. For example, using regression models, Singhal et al. [6] analyzed the impact of weather on New York City Transit subway ridership based on day of week and time of day combinations. They found that the impact of weather on transit ridership varies based on the time period and the location of subway stations.

Many machine learning algorithms were developed to model transit ridership. Liu and Chen [8] developed a deep learning method to predict the ridership at four BRT stations in Xiamen, China. They used a three-stage hybrid deep network model centered on hourly basis [8]. Also, Liu et al. [9] proposed a multilayer deep-learning architecture to predict metro inbound/outbound passenger flow. Other machine learning models to predict subway station ridership were also developed including gradient boosting decision trees [10], support vector machine [11], network Kriging method [12], and more. On the other hand, time series models have also been utilized in transportation related problems as well as to model the subway ridership. For example, time series ARIMA models have been applied to many areas of transportation including traffic arrival demand modeling [32], seasonal variation of freeway traffic conditions [33], prediction of actuated signal cycle length [19], traffic speed modeling on a downstream link [34], etc. Also, time series models have been applied to predict transit ridership; for instance, Ding et al. [7] presented an ARIMA-GARCH time series model to predict short-term metro ridership, which is an ARIMA model that takes care of the deterministic part and the non-linear GARCH model for the stochastic part. Their proposed ARIMA-GARCH model was applied on three stations ridership in Beijing and the result showed that it outperformed other proposed models [7].

Many of the existing time series models in the field of transportation have focused on stationary or smoothly varying models. However, these models cannot be used when there is a dramatic change in the system that could affect the time series model. For example, when Hurricane Sandy occurred in New York City in 2012, a sudden change of 11% decline happened in the number of trains in service on an average day [28]. This major disruptive event could impact on the future growth of transit ridership. Other disruptions include transit strike, bridge closure, special events like Olympic Games, and earthquake which the impact of these events on transit ridership and mode choice were reviewed in the study by Zhu and Levinson [29]. For example, considering transit strike, a study of 13 cases between 1966 to 2000 in the United States and Europe showed that it caused short-term and long-term losses on transit ridership [30]. Events like transit strikes provide a unique opportunity to study the change in transit ridership and travelers' behaviors, both of which are significant for drafting future transit policies. Stationary statistical models are not appropriate in the presence of anomalies caused by an unforeseen event. Global pandemic can be considered as an unforeseen event that significantly violates the stationary assumption of time series. Change point detection methods could be a remedy to unravel this issue in time series models.

To find a change point detection, it is possible to heuristically approximate locations of break points and estimate parameters within each (heuristically) estimated stationary segment. However, there are three main advantages in using a data-driven method to perform break point detection. First, in general, it is not clear whether an external shock would necessarily yield to a discontinuity point in the model. Hence, one may think that a time point should be considered as a break point while there are no break points in the data. Second, an important assumption in change point detection analysis is that the total number of break points in *unknown* and must be estimated from data. Therefore, one should first get an estimate on the number of change points in the data, and then, try to approximate location of break point(s) using either heuristic methods or data-driven ones. However, estimating the number of change points in the data is a complicated task and there is no clear way how to perform it heuristically. Third, estimating location of break points is a complicated task for heuristic methods since there may be a delay from the time of external shock to real changes in the parameters in the data or it may be the case that model parameters start to change before the external shock is known to human. As a result, heuristic methods may estimate location of break points with certain error which may yield to inaccurate estimation of model parameters. Therefore, data-driven methods can estimate number of change points and their locations more accurate than heuristic methods.

Data-driven methods to perform change point detection have recently been applied to the transportation research domain. Tang and Gao developed a nonparametric model for traffic flow prediction utilizing anomaly detection [13], while Tsiligkaridis and Paschalidis [14] applied anomaly detection in order to detect traffic jams. Margarieter leveraged anomaly detection methods in her research to detect incidents [15]. Riverio et al. [16] presented an analytical framework to detect anomalous events to a large real road traffic dataset collected from various areas in Europe. Moreover, an anomaly detection based analytical module was developed by Zhang et al. [17] to visualize the abnormal passenger traffic flow on urban network. In this paper, the authors model the subway station ridership during the COVID-19 pandemic by considering specific time series models combined with possible change points. Next, this rich family of time series models is introduced.

## TIME SERIES MODELS

Time series models are very well-known statistical models designed for data sets indexed by time. Time series models have been employed to different disciplines including finance, water resources, climate change, transportation, etc. [19][20][21]. The main objective of time series models is to grasp the underlying behavior of the dataset over time, and then, to detect the dependence among such data points in order to forecast future. More specifically, in these models, one tries to find the linear dependency structure among the data points over time, and then use that dependency to predict the new points in the future. This research focuses on applying Auto Regressive Integrated Moving Average (ARIMA) models. ARIMA model is a well-established yet simple statistical tool to deal with data indexed by time [22]. In the next section, a short introduction of this family of models is provided.

### *ARIMA MODEL*

Auto Regressive Integrated Moving Average (ARIMA) model is a powerful statistical tool when one is dealing with univariate time series or one data set. Suppose there is a data set $X_1, X_2, \ldots, X_n$, which is observed through time. For example, $X_1$ is the observation at the first time point, $X_2$ is the observation at the second time point, and so on. The Auto Regressive Moving Average ($ARMA$) model assumes that the present value of a time series is a linear combination of its past observations together with a linear combination of noises in the past observations. Thus, the time series $X_t$ is called $ARMA(p, q)$ as follows:

$$X_t - \phi_1 X_{t-1} - \cdots - \phi_p X_{t-p} = Z_t + \theta_1 Z_{t-1} + \cdots + \theta_q Z_{t-q} \qquad (1)$$

Here $\phi_1, \phi_2, \ldots, \phi_p$ are defined as $AR$ constants, $\theta_1, \theta_2, \ldots, \theta_q$ are defined as moving average or $MA$ constants, and $Z_t$ is called white noise (WN) with mean 0 and variance $\sigma^2$ ($WN(0, \sigma^2)$). In the stated model, the current value of data point stems from the past $p$ observations through $\phi_i$'s and the past $q$ observation noises through $\theta_i$'s. ARMA models are stationary models. In stationary models, the covariance between two observations $X_t$ and $X_{t+h}$ depends only on the lag $h$ and not on the time $t$. To put it simply, in stationary model, the dependence structure of points exits on their distances and not on their locations [22]. ARIMA models are an extension of ARMA models to capture the non-stationary behavior in the model. $ARIMA$ models are generally denoted by $ARIMA(p, d, q)$, where p is the order (number of time lags) of the autoregressive part, d is the degree of differencing (the difference between the current observations and $d$ time lags in the past is calculated as the new data set), and q is the order for the moving-average part. Noteworthy, the $ARIMA(p, 0, q)$ is also an $ARMA(p, q)$ model; $ARIMA(p, d, q)$ will be non-stationery when $d \neq 0$.

To measure the degrees of dependency among data points at different times, an autocorrelation function called ACF is used. ACF at lag h is calculated as

$$\rho(h) = \frac{\gamma(h)}{\gamma(0)} \qquad (2)$$

Where the $\gamma(h)$ is the autocovariance function (ACVF) at lag $h$ and is defined as:

$$\gamma(h) = Cov\ (X_t, X_{t+h}), \tag{3}$$

where $Cov$ means the covariance of two datasets. The $ACF$ at lag $h$ is the normalization version of $ACVF$. The partial auto-correlation function ($PACF$) provides the partial correlation of a time series with its own lagged values, controlling for the values of the time series in between them. For the time series $X_t$, the $PACF$ at lag $h$ denoted by $α(h)$ is the auto-correlation between $X_t$ and $X_{t+h}$ given the points $X_{t+1},…, X_{t+h-1}$. $ACF$ and $PACF$ are important factors in estimating ARIMA model's parameters, *p, d, q,* which will be used to predict the future values.

The main assumption for parameters in ARIMA models is that the roots of the auto-regressive polynomial should not be on the unit circle, i.e., should not have complex norm of one. Indeed, all estimated ARIMA models satisfy this property. This ensures that each model is stationary/stable. Moreover, to ensure the model is causal (invertible), all roots of the auto-regressive polynomial (moving average polynomial) should be outside of the unit disc, i.e., should have complex norm greater than one (see more details in [22]). This assumption for the case of p=1 or q=1, i.e., AR(1) or MA(1) model is equivalent to the phi/theta parameter being less than one. However, for higher lags (p or q more than 1), it is possible to have phi or theta parameters which have magnitudes more than one.

A piece-wise stationary ARIMA consists of several different and independent ARIMA models concatenated at certain time points called change points (or break points). In general, the number of change points and their locations are unknown and need to be estimated using statistical techniques. In the next section, a brief description of one such procedure to detect change points is provided.

*CHANGE POINT DETECTION*

Change point detection is an active line of research in statistics, specifically in the field of time series analysis with applications in many scientific fields such as Economics and Health Sciences. The main objective of this line of research is to find time point(s) at which the parameters of data generating process have changed. This change may be in the mean, variance, covariance structure or the spectral density of the time series. In this paper, we leverage the procedure developed in Safikhani and Shojaie [23] to detect the time when the change or break point occurs in the data sets under investigation.

The detection algorithm developed in Safikhani and Shojaie [23] has three main steps while the first two are for detection purposes and the last step is for model parameter estimation within each stationary segment. The first step assumes every time point is a (potential) break point. This is done by expanding the model parameter space. Then, the fused lasso penalty combined with the least squares objective function is utilized to find a set of candidates change points by reformulating the detection problem as a variable selection problem in a high-dimensional regression model. This step ensures that no true break point remains isolated, i.e., there will be at least one estimated break point close to any true break point with high probability (under certain regularity conditions). On the other hand, the total number of candidates change points selected in the first step may be larger than the true number of break points. To that end, a screening step is added (step 2) to search over all candidate change points and only keep the ones which would reduce the combined mean squared error (MSE) on the left- and right-hand side of the estimated break point significantly compared to MSE by ignoring the estimated break point. This screening step is shown to remove the redundant break points estimated in the first step with high probability, hence the final estimated change points are consistent estimates for location of break points and the number of estimated change points is a consistent estimate for the true number of change points asymptotically. Finally, the third step estimates the model parameters in all stationary segments by applying penalized estimation techniques using lasso in high dimensions (or the simple least squares method in low dimensions).

The developed algorithm in Safikhani and Shojaie [23] works for general multivariate vector auto-regressive models, thus a perfect fit to our modeling framework. This algorithm can handle detection for auto-correlation (the second order structure). The basic idea is to first assume every time point could potentially be a break point and fit a smooth version of time-varying parameters to the data. This step is performed using the fused lasso technique by Rinaldo [24]. Specifically, the time series model can be written as a linear regression model with very large design matrix as

$$Y = X\boldsymbol{B} + E, \tag{4}$$

where $Y$ is the vector of currently observed data points, X is a lower-triangular design matrix consisting of past values of time series (up to p lags), and E is the usual measurement error which is generally assumed to be independent and identically distributed (i.i.d). while this assumption can be relaxed (see e.g. [23]). The unknown parameter vector **B** can be estimated by minimizing a double-regularized least squares:

$$\widehat{\mathbf{B}} = argmin_{\mathbf{B}} \frac{1}{n}\|Y - Z\mathbf{B}\|_2^2 + \lambda_{1,n}\|\mathbf{B}\|_1 + \lambda_{2,n}\sum_{k=1}^{n}\left\|\sum_{j=1}^{k}\beta_j\right\|_1. \quad (5)$$

In equation (5), the notation $\| \ \|_2$ is the Euclidean norm of a vector while $\| \ \|_1$ is the sum of absolute values of elements in a vector. Further, n is the number of time points in the data, $\beta_j$ is the jump parameter at time point j, and $\sum_{j=1}^{k}\beta_j$ is the model parameter at time point k. Finally, the tuning parameters $\lambda_{1,n}$ and $\lambda_{2,n}$ are non-negative sequences of real numbers which are selected using data-driven techniques. The first term in the objective function in equation (5) is the mean squared error term, followed by two $l_1$ penalties controlling the number of break points and the sparsity of the auto-regressive parameters. The first one (i.e. $\lambda_{1,n}\|\mathbf{B}\|_1$) is the main component since it fuses the differences between estimated parameters to ensure smoothly estimated parameters over time, and it's called fused lasso [25][27]. The second one (i.e. $\lambda_{2,n}\sum_{k=1}^{n}\left\|\sum_{j=1}^{k}\beta_j\right\|_1$) is needed only in the high-dimensional regime where the number of time series components are much larger than the number of time points at which the data was observed. The two tuning parameters $\lambda_{1,n}$ and $\lambda_{2,n}$ can be approximated using rolling-window type cross-validations [17]. Specifically, since we focus on univariate modeling in this paper, we can simply put $\lambda_{2,n}=0$ which does not violate the assumptions in [23]. The optimization problem (5) is convex and can be solved efficiently. In fact, by Proposition 1 of the Friedman et al. [27], problem (5) can be solved by first finding a solution $\widehat{\mathbf{B}}^{(0)}$ for $\lambda_{2,n} = 0$ and then applying element-wise soft-thresholding to $\widehat{\mathbf{B}}^{(0)}$ in order to obtain the final estimate $\widehat{\mathbf{B}}$ for $\lambda_{2,n}\neq 0$. We refer to [23] for more details on the optimization problem (5) and how to select the tuning parameters in finite sample.

An interesting fact about the first step (fused lasso) is that this step might lead to over-estimation of number of break/change points in the model. Thus, in the second step, the redundant candidate change points detected in the first step are removed through a careful screening step. We omit the details on the second step and refer to [23], Section 4, equations (9-12) for more details on the screening step. Under basic regularity conditions and enough jumps in the parameters, this procedure is proved to consistently estimate the number of change points and their locations [24]. In the third and final step, model parameter estimates in each segment in which there are no more change points are estimated using the least squares ideas combined with penalization in the high-dimensional regime (similar to the estimation procedure (5)).

# DATA PREPARATION

The data used for the model is the subway station ridership from January 2019 to July 2020, retrieved from the New York City Transit open data system. New York City Subway system is operated by the Metropolitan Transportation Authority (MTA). It includes 472 stations in operation and approximately 4,600 turnstiles. The subway station turnstile ridership, either entries or exits, are provided on a four-hour increment by the MTA. In this research, the authors picked some subway stations in different boroughs in NYC. The selection of subway station was made randomly while making sure that there are no inconsistencies in the data during the study period. The busiest subway stations functioning as transport hubs were of limited use for the purpose of this research because they include many subway lines that might not work all the time, as they do not follow the data continuity of the time series.

In order to perform statistical analysis as well as anomaly detection, a data preparation is done on some subway stations in New York City. The authors look first at the data for South Ferry Subway station located in Lower Manhattan. It is located just close to the Ferry station from the borough of Manhattan to the borough of Staten Island, and it is also near the City's financial district, home to Wall Street and glittering skyscrapers. The data was aggregated to a daily basis from subway turnstile entries. During data processing and cleaning, some errors were noticed regarding turnstile records.

In the preprocessing part of calculating the entry ridership, a data warehouse was created using the MTA turnstile data. In that process, three categories were created which include: (i) subunit channel position (SPC), which represents a specific address of the used device; (ii) date of the associated data point; (iii) time of the associated datapoint. There is a metric that shows the cumulative entry of registered value of an associated device. These cumulative entries were subtracted from the ones of the previous days, to give the daily "entries". The MTA records subway ridership at every turnstile, which basically report both entry and exit. The number of ridership reported, either entries or exits, are turnstile counter values. When the counter reaches its maximum limit, the counter is reset. A drastic shift in the absolute (cumulative) number reported in the data is noticed during the timer period when the reset happens. This issue is accounted in the data cleaning process. To do this, when the cumulative sequence value changed significantly, the rolling average of the "entries" subway values was calculated.

To understand the trends of subway ridership, the ridership data over time at the South Ferry Subway station is used for illustration. Figure 1 shows the daily turnstile entries and moving average of the mean over a seven-day rolling window at the South Ferry Subway station, starting from January 1st in 2020 to June 19th in 2020. This ridership includes all turnstile entries of the South Ferry Subway station. The rolling average method reduces noise in time series data, enabling the ability to look at obvious trends. The subway ridership decreased gradually, then plunged suddenly in March 2020, and continued to perform with a very low ridership during April 2020 as shelter-in-place orders were mandated by state officials.

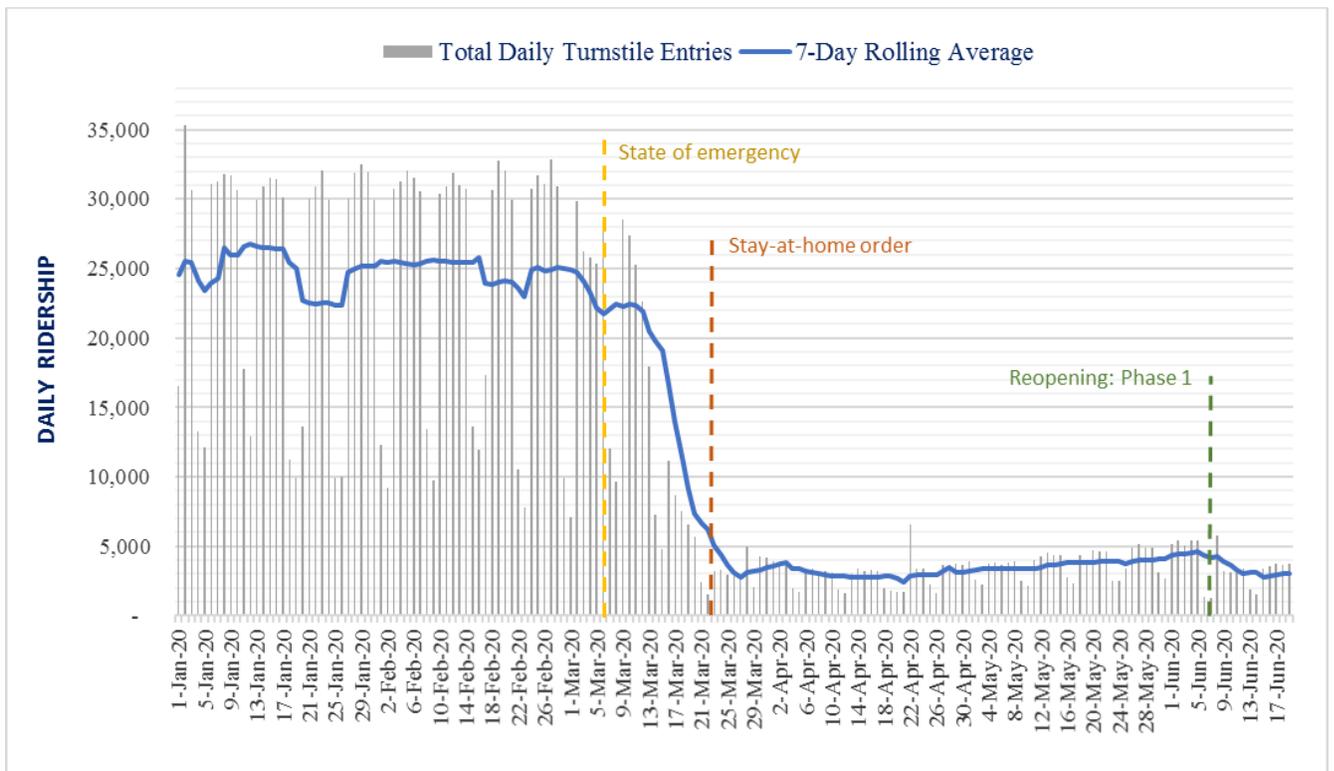

**FIGURE 1: Daily entries and 7-day rolling average of the mean of South Ferry Subway station**

In the following, more descriptive analyses of this subway ridership are provided. Graphs of Figure 2 show the histogram, autocorrelation function (ACF), partial autocorrelation function (PACF) of the raw ridership at the South Ferry Subway station. The histogram is skewed to the left as a portion of dataset includes a time period during the COVID-19 pandemic which imposed a low ridership to subway. From ACF and PACF plots, it is noticeable that there is a seasonal pattern in the data. This seasonality is mostly due to the weekly mobility of commuters utilizing subway for work purposes on weekdays.

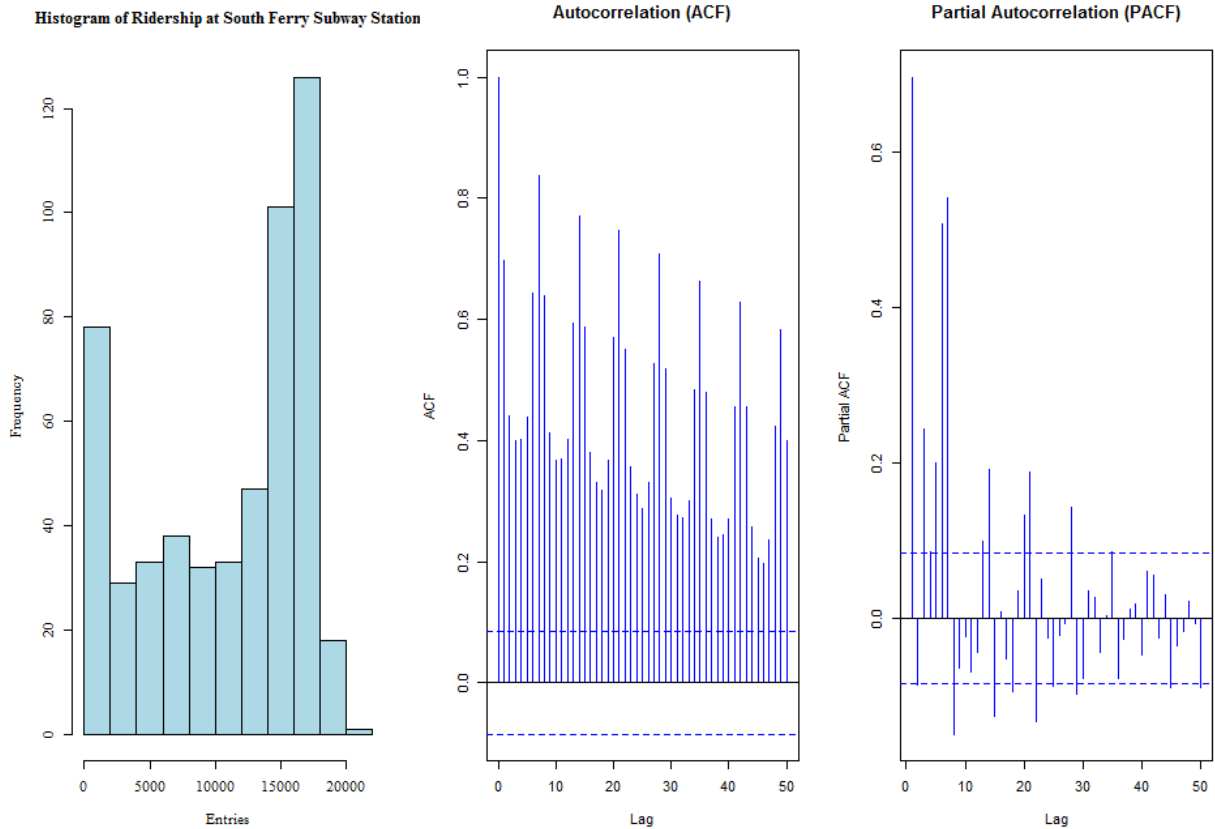

**FIGURE 2: Histogram, autocorrelation function (ACF), and partial autocorrelation function (PACF) of the turnstile entries raw data at South Ferry Subway station, from Jan 2019 to April 17th, 2020**

To better understand the patterns between different data points in time series, the time series' trend and seasonality are removed from the dataset. Then, one will just focus on the time series of the residuals in order to decipher any temporal dependencies among them. First, a linear regression was applied to incorporate the imbalance between weekdays and weekends ridership. To do so, Saturday and Sunday (the weekend days) are labeled as 1, other weekdays are considered as 0. In the linear regression model, the dummy categorial weekday/weekend indicator, being 0 as weekday and 1 as weekend, is defined as the independent variable. After applying the linear regression, the residual from the regression model is left. Second, to remove seasonality in our dataset, a seasonal autoregressive moving average model (SARIMA) was applied with periodic operator of 7. It treats the dataset as if there is a weekly seasonal trend. Finally, the residual from the SARIMA model, which is the absolute error of the ridership, is the output to be analyzed for the rest of this paper. It is worth noting that no transformation, i.e. log-transform, was applied to the raw dataset. Figure 3 shows the time series of residual after removing the trend and seasonality. The vertical green, blue, and red lines in the graph correspond to January 1$^{st}$, March 22$^{nd}$ (the day for state-at-home order), and June 8$^{th}$ (the day for reopening of Phase 1) in the year 2020, respectively. Also, Figure 4 displays time series trends of all seven subway stations after removing the trend and seasonality.

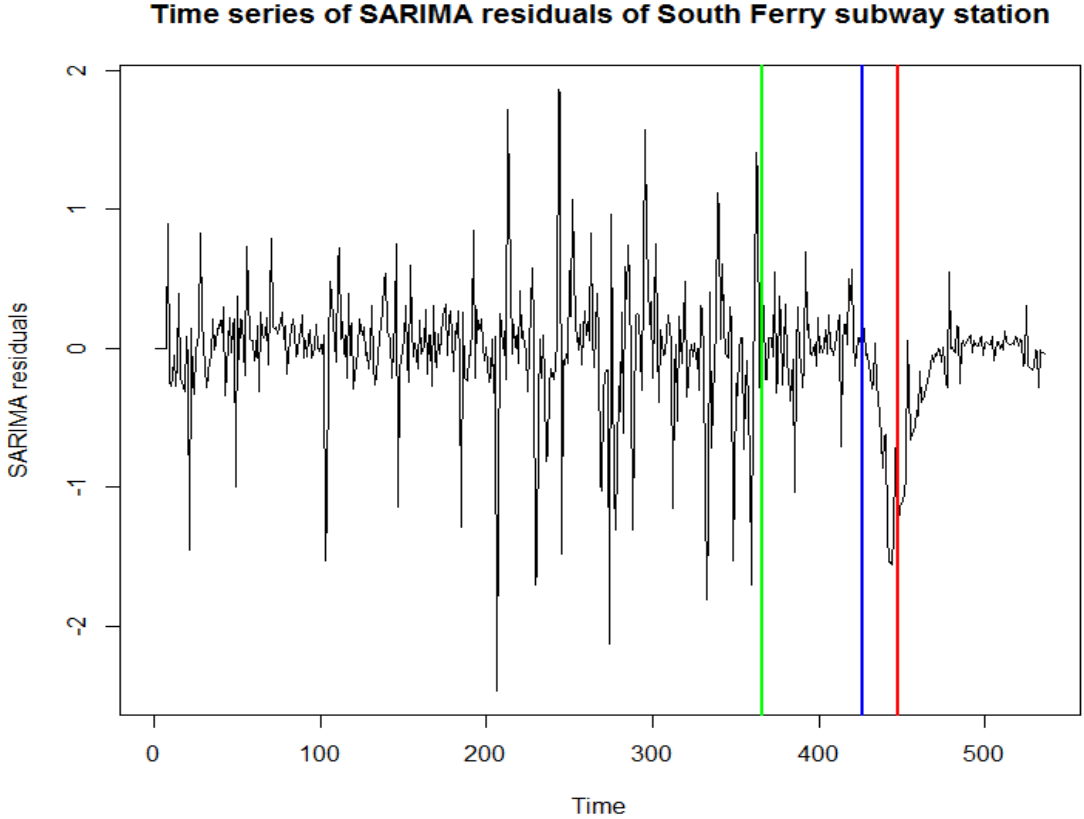

**FIGURE 3: Residuals of the SARIMA model (South Ferry Subway station)**

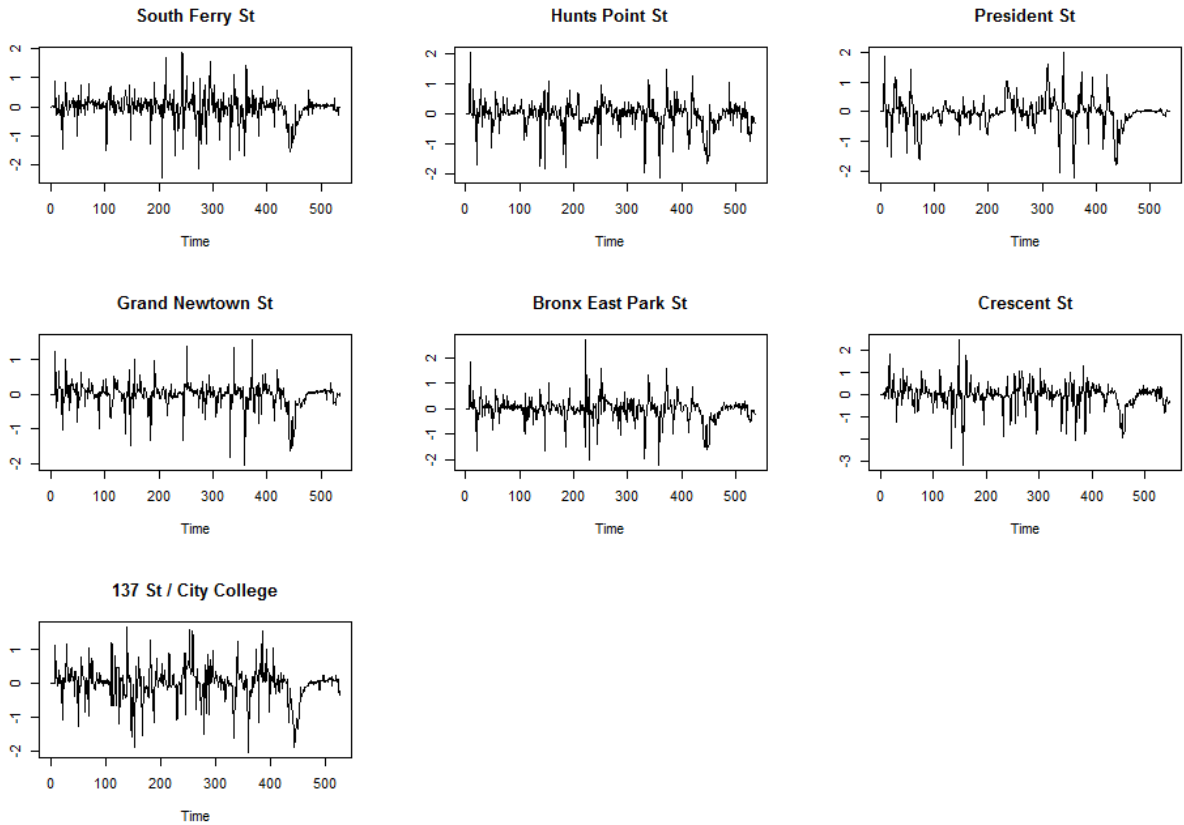

**FIGURE 4: Residuals of the SARIMA model for all seven subway stations**

Figure 5 illustrates the ACFs of the residuals at every season from the beginning of 2019 to June 2020. It shows how the autocorrelation changes from one season to another. All four ACF plots in 2019 have the same shape, mostly significant at lag 1 and lag 7 with a different magnitude. However, the ACF for the winter season in 2020 shows a different pattern when it is compared to other seasons, specifically to the corresponding seasons in 2019. Specifically, there is a strong autocorrelation among data points in the winter season 2020 compared to the same time in 2019. Such extreme temporal correlations resemble the existence of change points in the data which appears as long-range dependence [25]. Similar pattern has been observed for other subway stations considered here. This observation justifies the use of piece-wise stationary time series modeling framework for the data sets focused on this project.

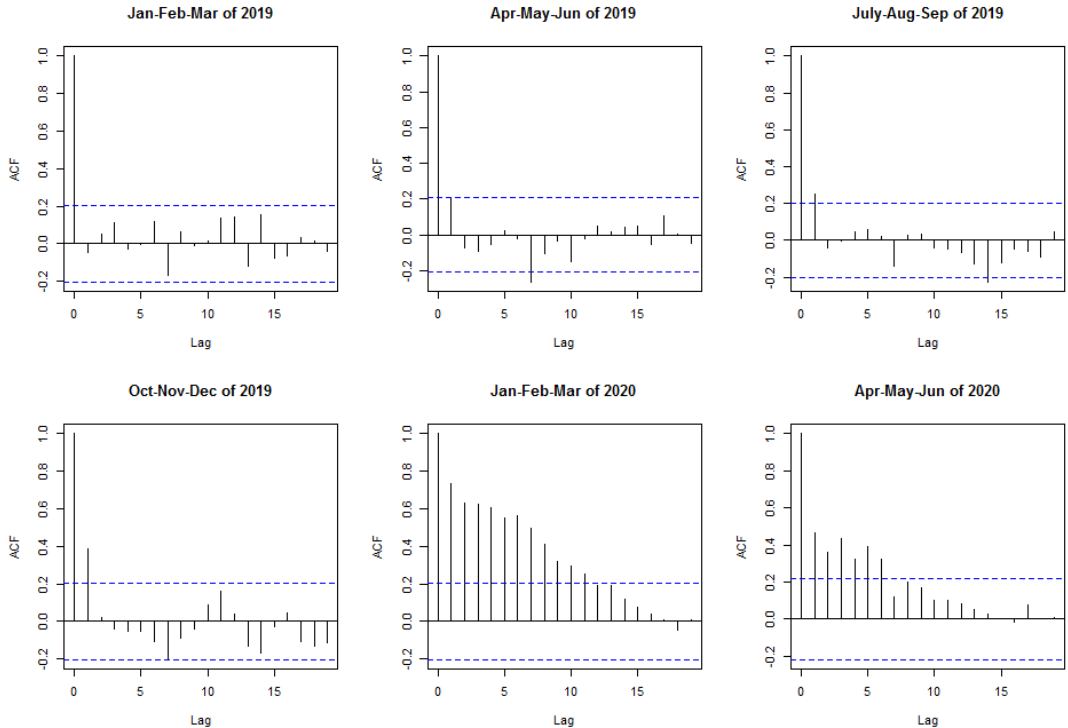

**FIGURE 5: ACFs of each season from beginning of 2019 until June 2020 (South Ferry Subway station)**

Since the ACF of winter season in 2020 is different, this season has been divided into three segments resembling the three months of winter. Figure 6 displays the ACFs of the first three months of 2020 (January, February, and March) which illustrates how the autocorrelation changes from one to another. One can noticeably observe that the ACF in March has a different behavior. March 2020 corresponds to the time when the positive coronavirus cases emerged and increased to the point when New York state officials announced the state of emergency, then the New York State's stay-at-home order took effect [3].

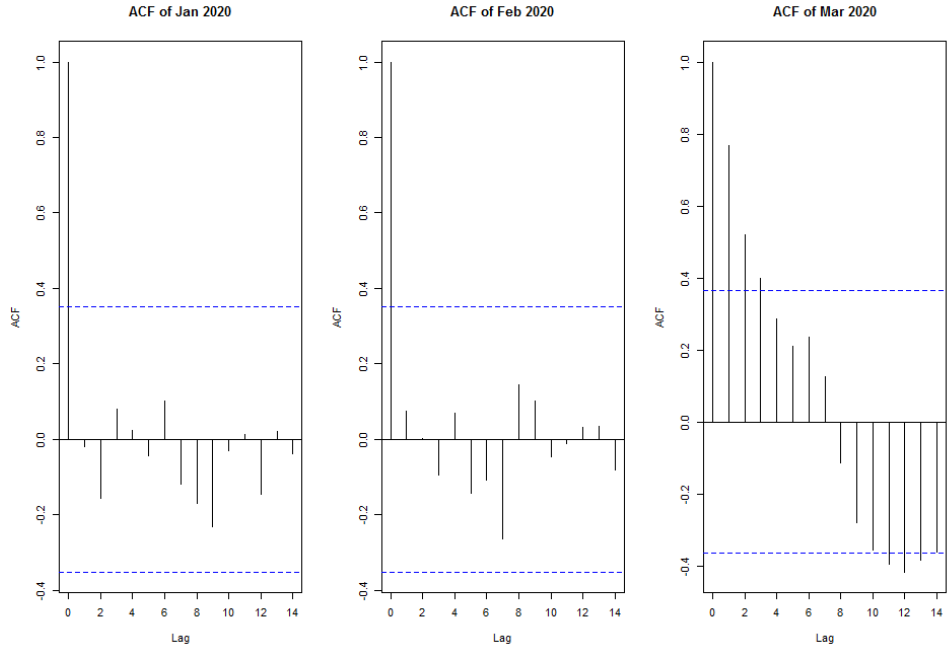

**FIGURE 6: ACFs of three months: January, February, and March in 2020 (South Ferry station)**

One of the main goals of this paper is to understand the subway's demand variability during unexpected events, in the face of COVID-19 pandemic. As explained before, the approach is to consider existence of anomalies at which the subway demand has suddenly changed, either surged or plunged. The existence of such anomalies in subway ridership would violate the assumption of stationarity in the time series models which is an underlying assumption in many non-stationary times series models in the literatures.

In the first attempt of modeling, we ignore the existence of change points and simply apply ARIMA models by applying the function "auto.arima" in the R programming language to the processed data. Table 1 illustrates the model summary of applying ARIMA model to the processed data. The table includes model's name and its estimated parameters, model's performance errors by Akaike information criterion (AIC), corrected AIC (AICC), and Bayesian information criterion (BIC), as well as the p-values from the white noise Li-McLeod's test [22]. The first row in the table describes the summary of the fitted ARIMA model to the whole data. The fitted model for the whole data is ARIMA (0,1,2), which gives the best ARIMA model based on AICC. However, it turns out that this model is not a good fit for the data since the residuals turn out not to be white noise.

**TABLE 1: Time series models for the whole data of the South Ferry Subway station ridership**

| Data | Fitted Model | Estimated Parameters | Performance Errors | Li-McLeod's test P-values |
|---|---|---|---|---|
| **Whole Data: Jan 2019 to June 2020** | ARIMA (0,1,2) | $\theta_1 = -0.6368$<br>$\theta_2 = -0.2590$ | AIC = 653.81<br>AICC = 653.85<br>BIC = 666.65 | 0.012 (less than 0.05) |

The white noise test was performed using Li-McLeod test method (see [22] for further information about the white noise test). The *p*-value of the white noise test for the fitted mode was equal to 0.012, which indicates that there are still some dependencies exist among lagged data points which requires further modeling and modification.

The same simple analysis was performed on the subway ridership for other randomly selected subway stations in New York City. Specifically, seven subway stations that have been chosen randomly across different boroughs are: South Ferry Subway station in lower Manhattan, Crescent Street in Brooklyn, Bronx Park East station in Bronx, 137 Street (the City College) station in Upper Manhattan, Hunts Point Avenue station in Bronx, President Street station in Brooklyn, and Grand Avenue-Newtown station in Queens. The subway stations were selected so that each time series of ridership does not to have any missing data. Table 2 shows the fitted ARIMA models at all seven selected subway stations with their coefficient values.

**TABLE 2: Time series model for all four subway stations**

| Subway Station | Fitted Model | Estimated Parameters | Performance Errors | Li-McLeod's test P-values |
|---|---|---|---|---|
| **South Ferry station** | ARIMA (0,1,2) | $\theta_1 = -0.6368$<br>$\theta_2 = -0.2590$ | AIC= 653.81<br>AICC= 653.85<br>BIC= 666.65 | < 0.05 |
| **137 Street- City College** | ARIMA (1,0,2) | $\phi_1 = 0.886$<br>$\theta_1 = -0.342$<br>$\theta_2 = -0.299$ | AIC = 610.36<br>AICC = 610.44<br>BIC = 627.42 | < 0.05 |
| **Bronx East Park station** | ARIMA (1,1,1) | $\phi_1 = 0.219$<br>$\theta_1 = -0.871$ | AIC = 587.61<br>AICC = 587.65<br>BIC = 600.45 | < 0.05 |
| **Crescent Street - Brooklyn** | ARIMA (1,0,1) | $\phi_1 = 0.435$<br>$\theta_1 = 0.148$ | AIC = 754.63<br>AICC = 754.58<br>BIC = 767.54 | < 0.05 |
| **Hunts Point station** | ARIMA (1, 0, 2) | $\phi_1 = 0.819$<br>$\theta_1 = -0.397$<br>$\theta_2 = -0.145$ | AIC= 527.04<br>AICC= 527.12<br>BIC= 544.18 | < 0.05 |
| **President Station** | ARIMA (1, 0, 0) | $\phi_1 = 0.682$ | AIC= 390.7<br>AICC= 390.72<br>BIC= 399.27 | < 0.05 |
| **Grand Newton station** | ARIMA (2, 1, 4) | $\phi_1 = -0.4081$<br>$\phi_2 = -0.7528$<br>$\theta_1 = -0.1235$<br>$\theta_2 = 0.3702$<br>$\theta_3 = -0.6168$<br>$\theta_4 = -0.2732$ | AIC= 311.76<br>AICC= 311.97<br>BIC= 341.74 | < 0.05 |

Existence of long-range dependence in the data together with failing to model the processed data sets as stationary ARIMA model motivated the use of ARIMA modeling with change points. The first step to develop such a model is to find change points in each data set.

After running the change point detection algorithm developed in [23], the results of each subway station data are reported in the following Table 3. The change point for the 137 Street station (the City College) in Upper Manhattan is March 13[th], 2020 while the change point for Bronx East Park station is March 12[th], 2020. Further, the change point for Crescent Street station in Brooklyn is March 13[th], 2020; the Change point for South Ferry Subway station in Lower Manhattan is March 13[th], 2020. Two change points detected for Hunts Point station are June 28[th], 2019, and March 12[th], 2020. Also, the change point for President Station and Grand Newton stations is detected to be March 13[th], 2020.

TABLE 3: Change Point of subway station ridership

| Subway Station | Change Point (Time) |
|---|---|
| **137 Street- City College** | March 13[th], 2020 |
| **Bronx East Park station** | March 12[th], 2020 |
| **Crescent Street -Brooklyn** | March 13[th], 2020 |
| **South Ferry station** | March 13[th], 2020 |
| **Hunts Point station** | June 28[th], 2019 & March 12[th], 2020 |
| **President station** | March 13[th] -2020 |
| **Grand Newton station** | March 13[th], 2020 |

Understandably, all change points detected for the subway station ridership considered are within almost a week from the state of emergency reported in the state of New York. Detecting another change point for the Hunts Point station around June 28th in 2019 is related to maintenance and partial closure of this subway station which caused a significant drop in its ridership. Such a consistent performance of the detection algorithm on all subway stations considered here is an indication that the piece-wise stationary time series model focused on this project is a reasonable choice. Next, we apply ARIMA models to each stationary segment estimated by our method in order to further investigate the goodness-of-fit for our proposed model.

With change points values being found, instead of decomposing the data through eyeballing procedure, the subway ridership data at each station has been separated into before and after the change point(s). Table 4 shows the fitted ARIMA model (again, applying the function "auto.arima" in the R programming language) for each subway station before and after the change point(s). It should be noted again that before running the change points algorithm, the seasonality and trend of each dataset were already considered.

**TABLE 4: Time series model for all four subway stations before and after change point**

| Subway Station | Fitted Model | Estimated Parameters | Performance Errors | Li-McLeod's test P-values |
|---|---|---|---|---|
| **South Ferry station (before change point)** | ARIMA (0,0,1) | $\theta_2 = 0.2677$ | AIC= 563.07<br>AICC= 563.1<br>BIC= 571.23 | >= 0.05 |
| **South Ferry station (after change point)** | ARIMA (3,1,3) | $\phi_1 = 1.3867$<br>$\phi_2 = -1.0280$<br>$\phi_3 = 0.0795$<br>$\theta_1 = -1.9053$<br>$\theta_2 = 1.5240$<br>$\theta_3 = -0.370$ | AIC=-34.28<br>AICC=-33.03<br>BIC=-16.26 | >= 0.05 |
| **137 Street- City College (before change point)** | ARIMA (1,0,1) | $\phi_1 = 0.310$<br>$\theta_1 = 0.492$ | AIC = 486.4<br>AICC = 486.5<br>BIC = 498.7 | < 0.05 |
| **137 Street- City College (after change point)** | ARIMA (5,0,1) | $\phi_1 = 0.0536$<br>$\phi_2 = -0.8238$<br>$\phi_3 = 0.4975$<br>$\phi_4 = -0.1208$<br>$\phi_5 = 0.2307$<br>$\theta_1 = 0.4355$ | AIC = 531.2<br>AICC = 531.53<br>BIC = 563.84 | >= 0.05 |
| **Bronx East Park station (before change point)** | ARIMA (0,0,2) | $\theta_1 = 0.1761$<br>$\theta_2 = 0.1947$ | AIC = 491.35<br>AICC = 491.4<br>BIC = 503.58 | >= 0.05 |
| **Bronx East Park station (after change point)** | ARIMA (2,1,1) | $\phi_1 = 0.7612$<br>$\phi_2 = -0.1742$<br>$\theta_1 = -0.8191$ | AIC = -15.55<br>AICC = -15.12<br>BIC = -5.25 | >= 0.05 |

| Crescent Street - Brooklyn (before change point) | ARIMA (0,0,2) | $\theta_1$= 0.5124<br>$\theta_2$= 0.1316 | AIC = 659.27<br>AICC = 659.32<br>BIC = 671.58 | >= 0.05 |
|---|---|---|---|---|
| Crescent Street - Brooklyn (after change point) | ARIMA (0,1,0) | | AIC = 4.83<br>AICC = 4.87<br>BIC = 7.42 | >= 0.05 |
| Hunts Point (before) | ARIMA (0,0,1) | $\theta_1$= 0.2199 | AIC= 178.2<br>AICC= 178.27<br>BIC= 184.57 | >= 0.05 |
| Hunts Point (after, 2 models) | ARIMA (3,0,0) | $\phi_1$= 0.4740<br>$\phi_2$=-0.0671<br>$\phi_3$= 0.0522 | AIC= 262.91<br>AICC= 263.07<br>BIC= 277.14 | >= 0.05 |
| | ARIMA (2,1,1) | $\phi_1$= 0.3942<br>$\phi_2$= 0.2132<br>$\theta_1$= -0.9014 | AIC= 64.67<br>AICC = 65.1<br>BIC= 74.96 | >= 0.05 |
| President station (before) | ARIMA (1,0,0) | $\theta_1$= 0.6636 | AIC= 394.15<br>AICC= 394.18<br>BIC= 402.31 | >= 0.05 |
| President station (after) | ARIMA (6,1,0) | $\phi_1$= 0.1121<br>$\phi_2$= -0.6223<br>$\phi_3$= -0.3153<br>$\phi_4$= -0.1176<br>$\phi_5$= -0.5350<br>$\phi_6$= 0.3523 | AIC= -140.1<br>AICC= -138.48<br>BIC= -119.42 | >= 0.05 |
| Grand Newton station (before) | ARIMA (1,0,0) | $\phi_1$= 0.3244 | AIC=270.06<br>AICC=270.09<br>BIC=278.23 | < 0.05 |
| Grand Newton station (after) | ARIMA (2,1,2) | $\phi_1$= 1.3181<br>$\phi_2$= -0.9618<br>$\theta_1$=-1.4192<br>$\theta_2$= 0.8813 | AIC=-52.96<br>AICC=-52.3<br>BIC=-40.09 | < 0.05 |

The results indicate that with the analyzed data, there are two models (before and after the change point) that could be very well developed to represent the variability in the data set. The two models (or three with multiple change points) have performed better compared to using only one model. To perform a goodness-of-fit, all residuals are tested using the same procedure as before, i.e. the Li-McLeod test method [22], and most of the computed p-values are greater than 5% which indicate that the new residuals don't resemble any additional temporal dependence, hence the piece-wise ARIMA model fitted to the data is a reasonable fit. It should be noted that in ARIMA models, for higher lags (p or q more than 1), it is possible to have phi or theta parameters with magnitudes more than one, which is the case for models in the stations at South Ferry and at Grand Newton.

Another interesting finding is the changes occurred in the ridership data are not only on means (average of ridership), but the second order statistics are changed as well. Reviewing the fitted models in Table 2 and Table 4 show that the order of ARIMA models (selected p, d, q) and the estimated auto-regressive and moving average parameters in different segments for each subway station are different. For example, the selected model for South Ferry data before the break point is a simple MA(1) while after the break point is an ARIMA(3,1,3). The former one is a very simple model with minor correlations among data points with one time-lag apart from each other while the latter is a more complicated model with 6 parameters, and the temporal dependence is much stronger than the first model. Therefore, the correlations among time-lagged ridership have changed during the pandemic. Such second order changes are not easy to detect by eyeballing or just plotting the data as a time series. This shows the necessity of developing statistical modeling frameworks which would include anomalies in the data in order to analyze and understand better the behavior of subway ridership over time during such extreme times (pandemic).

Finally, looking at the results in the Grand Newton station, the white noise test p-values are relatively low which means the current model may not be a good fit for this data. A non-linear time series model be a better choice for this specific station.

**CONCLUSION**

In this paper, a linear and non-stationary ARIMA-based time series model is developed in order to understand the temporal pattern of station-based subway ridership during the COVID-19 pandemic. The main objective of the paper is to propose an interpretable modeling framework to characterize the temporal pattern of subway ridership during the COVID-19 pandemic.

Interpretability lies in the modeling framework utilized in this manuscript. It is a common practice to model data with temporal index using stationary time series models, and ARIMA models are among the most well-known stationary time series models used in different scientific fields including civil engineering and transportation. However, in the presence of structural breaks (shocks) in the temporal dynamical system under consideration, the stationarity assumption may be violated; thus, one needs to search for alternative modeling frameworks. Nonparametric statistical models using B-splines and wavelet basis functions are models which may be able to handle the changes in the dynamical system. However, they are not easy to interpret. This is one of the drawbacks of nonparametric methods while they are useful in practice. Piecewise stationary models, on the other hand, are interesting models which are easier to interpret since a time point at which the dynamical system receives an external shock can be called a "break point", and its location can be estimated using developed algorithms. Such time points are essentially the discontinuity points in the piecewise modeling framework. This is why we call the modeling framework an interpretable one. For example, our developed algorithm detects a break point in the ridership of South Ferry Station on March 13$^{th}$, 2020. This date could be linked to an external shock in the dynamical system of ridership at this station which is indeed close to the stay-at-home order at the state level in NY. Note that it is almost impossible to make such connections to real events using nonparametric models.

To that end, a simple yet powerful family of time series models are utilized (ARIMA models) while the inclusion of the sudden changes (break points) in the model accounts for discontinuity/anomalous behavior of subway ridership. A unique feature of the developed model is that it balances between model complexity and model interpretability. In other words, the statistical model designed for the time series data under investigation is deliberately selected to be simple and interpretable while it is a valid model (i.e. fits very well to the data). One may note that not only the detected break points found by the algorithm are near the "stay-at-home" orders in New York City which validates the proposed modeling framework, but also the model shows that the "covariance structure" of the time series have changed before and after the pandemic.

In summary, the main contributions of the paper include (a) developing a simple and interpretable statistical modeling framework to analyze the subway ridership in NYC during the COVD-19 pandemic; (b) utilizing novel machine learning algorithms in the statistical literature to detect the location of change/break points while the number of such break points are assumed to unknown; (c) enhancing the scientific community's understanding of changes in the subway ridership during such pandemics by shedding some light on the second order changes of subway ridership, a topic worth of further investigation.

It is worth noting that there are two different approaches of change point detection in the statistical and engineering literature: (1) online detection; (2) off-line detection [31]. In the latter, the whole time series data is given to the modeler and the objective is to find the set of break/change points while in the former, a streaming data is observed (new data points are observed continuously) and the objective is to raise an alarm as soon as an anomalous pattern is observed in the data. The change point model developed in this paper is off-line since we used the whole observed data and then used novel algorithms to locate all break points. Developing online detection algorithms needs different statistical treatments (mainly, likelihood ratio tests), different modeling framework (distributional assumption on the time series) and different algorithms. Further, certain assumptions must be put in order to define identifiable break/change points which may not hold for the subway ridership data in NYC during the COVID-19 pandemic. For these reasons, the authors focused on off-line change point detection and leave the online detection as a fruitful future research direction. Moreover, analyzing and comparing the effect of COVID-19 on transit ridership of other metropolitan cities would be in the continuation of the author's research to better justify the need of elaborated time series models. Also, how the transit ridership reacts during this pandemic with reopening strategies or how the subway ridership returns in the aftermath of COVID-19 crisis could add other anomalies to our time series modeling. These questions are few important research directions that the authors plan to pursue in the near future.

# References

[1]     Bonaccorsi, G., Pierri, F., Cinelli, M., Flori, A., Galeazzi, A., Porcelli, F., ... & Pammolli, F. (2020). Economic and social consequences of human mobility restrictions under COVID-19. *Proceedings of the National Academy of Sciences*, *117*(27), 15530-15535.

[2]     https://www.nytimes.com/2020/03/22/nyregion/Coronavirus-new-York-epicenter.html

[3]     https://www.nytimes.com/interactive/2020/us/coronavirus-stay-at-home-order.html

[4]     Mobility Trends in New York City During COVID-19 Pandemic: Analyses of transportation modes throughout June 2020, Report by University Transportation Research Center Region 2, New York, July 2020.

[5]     McNally, M. G. (2000). The four-step model. *Handbook of transport modelling*, *1*, 35-41.

[6]     Singhal, Abhishek, Camille Kamga, and Anil Yazici. "Impact of weather on urban transit ridership." *Transportation research part A: policy and practice* 69 (2014): 379-391.

[7]     Ding, C., Duan, J., Zhang, Y., Wu, X., & Yu, G. (2017). Using an ARIMA-GARCH modeling approach to improve subway short-term ridership forecasting accounting for dynamic volatility. *IEEE Transactions on Intelligent Transportation Systems*, *19*(4), 1054-1064.

[8]     Liu, Lijuan, and Rung-Ching Chen. "A novel passenger flow prediction model using deep learning methods." *Transportation Research Part C: Emerging Technologies* 84 (2017): 74-91.

[9]     Liu, Yang, Zhiyuan Liu, and Ruo Jia. "DeepPF: A deep learning based architecture for metro passenger flow prediction." *Transportation Research Part C: Emerging Technologies* 101 (2019): 18-34.

[10]    Ding, Chuan, et al. "Predicting short-term subway ridership and prioritizing its influential factors using gradient boosting decision trees." *Sustainability* 8.11 (2016): 1100.

[11]    Liu, Shasha, and Enjian Yao. "Holiday passenger flow forecasting based on the modified least-square support vector machine for the metro system." *Journal of Transportation Engineering, Part A: Systems* 143.2 (2017): 04016005.

[12]    Zhang, Dapeng, and Xiaokun Cara Wang. "Transit ridership estimation with network Kriging: A case study of Second Avenue Subway, NYC." *Journal of Transport Geography* 41 (2014): 107-115.

[13]    Tang, S., & Gao, H. (2005). Traffic-incident detection-algorithm based on nonparametric regression. IEEE Transactions on Intelligent Transportation Systems, 6(1), 38-42.

[14]    Tsiligkaridis, A., & Paschalidis, I. C. (2017, November). Anomaly detection in transportation networks using machine learning techniques. In 2017 IEEE MIT Undergraduate Research Technology Conference (URTC) (pp. 1-4). IEEE.

[15]    Margreiter, M. (2016). Automatic incident detection based on Bluetooth detection in northern Bavaria. Transportation research procedia, 15, 525-536.

[16]    Riveiro, M., Lebram, M., & Elmer, M. (2017). Anomaly detection for road traffic: A visual analytics framework. *IEEE Transactions on Intelligent Transportation Systems*, *18*(8), 2260-2270.

[17]    Zhang, Y., Shi, H., Zhou, F., Hu, Y., & Yin, B. (2020). Visual analysis method for abnormal passenger flow on urban metro network. *Journal of Visualization*, 1-18.

[18]    Ding, C., Wang, D., Ma, X., & Li, H. (2016). Predicting short-term subway ridership and prioritizing its influential factors using gradient boosting decision trees. Sustainability, 8(11), 1100.

[19]    Moghimi, B., Safikhani, A., Kamga, C., & Hao, W. (2018). Cycle-length prediction in actuated traffic-signal control using ARIMA model. Journal of Computing in Civil Engineering, 32(2), 04017083.

[20]    Moghimi, B., Safikhani, A., Kamga, C., Hao, W., & Ma, J. (2018). Short-term prediction of signal cycle on an arterial with actuated-uncoordinated control using sparse time series models. IEEE Transactions on Intelligent Transportation Systems, 20(8), 2976-2985.

[21]    Safikhani, A., Kamga, C., Mudigonda, S., Faghih, S. S., & Moghimi, B. (2018). Spatio-temporal modeling of yellow taxi demands in New York City using generalized STAR models. International Journal of Forecasting.

[22]    P.J. Brockwell, and R.A. Davis, "Introduction to time series and forecasting", Springer Science & Business Media, 2006.

[23]    Safikhani, A., & Shojaie, A. (2020). Joint structural break detection and parameter estimation in high-dimensional non-stationary var models. *Journal of the American Statistical Association*, 1-26.

[24]    Rinaldo, A. (2009). Properties and refinements of the fused lasso. *The Annals of Statistics*, *37*(5B), 2922-2952.

[25]    Aue, A., & Horváth, L. (2013). Structural breaks in time series. *Journal of Time Series Analysis*, *34*(1), 1-16.

[26]    Tibshirani, Robert, Saunders, Michael, Rosset, Saharon, Zhu, Ji, & Knight, Keith. 2005. Sparsity and smoothness via the fused lasso. Journal of the Royal Statistical Society: Series B (Statistical Methodology), 67(1), 91–108.

[27]    Friedman, Jerome, Hastie, Trevor, Höfling, Holger, Tibshirani, Robert, et al. 2007. Pathwise coordinate optimization. The Annals of Applied Statistics, 1(2), 302–332.

[28]    Paul, Julene, and Michael J. Smart. "The Hangover: Assessing Impact of Major Service Interruptions on Urban Rail Transit Ridership." Transportation Research Record 2648.1 (2017): 79-85.

[29]    Zhu, Shanjiang, and David M. Levinson. "Disruptions to transportation networks: a review." Network reliability in practice. Springer, New York, NY, 2012. 5-20.

[30]    van Exel, N. Job A., and Piet Rietveld. "Public transport strikes and traveller behaviour." Transport Policy 8.4 (2001): 237-246.

[31]    Csörgö, Miklós, and Lajos Horváth. Limit theorems in change-point analysis. Vol. 18. John Wiley & Sons Inc, 1997.

[32]     S. Barua, A. Das, and K.C. Roy, "Estimation of Traffic Arrival Pattern at Signalized Intersection Using ARIMA Model", International Journal of Computer Applications, no.128 (1). pp. 1-6, 2015.

[33]     B.M. Willams, and L.A. Hoel, "Modeling and forecasting vehicular traffic flow as a seasonal ARIMA process: Theoretical Basis and Empirical Results", Journal of Transportation Engineering, no.129(6), pp.664-672, 2003.

[34]     P. Duan, G. Mao, C. Zhang, and S. Wang, Rio, Brazil. "STARIMA-based Traffic Prediction with Time-varying Lags", IEEE 19th International Conference on Intelligent Transportation System (ITSC), 2016.